\documentclass[a4paper]{article}
\pdfoutput=1
\usepackage{jinstpub}
\usepackage{lineno}

\usepackage{siunitx}
    \DeclareSIUnit\atmosphere{atm}
    \DeclareSIUnit\year{year}
    \DeclareSIUnit\parsec{pc}
    \DeclareSIUnit\decade{dec}
    \DeclareSIUnit\dBm{dBm}
    \DeclareSIUnit\ph{ph}
    \DeclareSIUnit\ppm{ppm}
\usepackage{float}
\usepackage{booktabs}
\usepackage{caption}
\usepackage{subfig}
\usepackage{gensymb}
\usepackage{lscape}
\usepackage{longtable}
\usepackage{rotating}
\usepackage{tikz}
\usepackage{xcolor}
\usepackage{pgfplots}
\usepackage{listings}
\usepgfplotslibrary{units}
\pgfplotsset{compat=1.5}
\usetikzlibrary{arrows, calc, decorations,shapes.misc, pgfplots.groupplots}
\tikzset{snake it/.style={decorate, decoration={snake, post length=2mm}}}
\usepackage{circuitikz} 
\usepackage{pgfplots}
\usepackage{hyperref}
\usepackage{array}
\usepackage{amsmath,amsfonts}
\usepackage{algorithmic}
\usepackage{graphicx}
\usepackage{textcomp}
\usepackage{xcolor}
\usepackage{mhchem}

\title{\boldmath Xenon doping of Liquid Argon in ProtoDUNE Single Phase}

\author[1,2]{Niccol\`o Gallice}

\affiliation[1]{Universit\`a degli Studi di Milano, Dept. of Phsyics, via Celoria 16, Milano 20133, IT}
\affiliation[2]{INFN Sezione di Milano, via Celoria 16, Milano 20133, IT}

\emailAdd{niccolo.gallice@unimi.it}

\abstract{
The Deep Underground Neutrino Experiment (DUNE) will be the next generation long-baseline neutrino experiment. The far detector is designed as a complex of four LAr-TPC (Liquid Argon Time Projection Chamber) modules with \SI{17}{\kilo\tonne} of liquid argon each. The development and validation of the first far detector technology is pursued through ProtoDUNE Single Phase (ProtoDUNE-SP), a \SI{770}{\tonne} LAr-TPC at CERN Neutrino Platform. 
Crucial in DUNE is the Photon Detection System that will ensure the trigger of non-beam events --- proton decay, supernova neutrino burst and BSM searches --- and will improve the timing and calorimetry for neutrino beam events.
Doping liquid argon with xenon is a known technique to shift the light emitted by argon (\SI{128}{\nano\meter}) to a longer wavelength (\SI{178}{\nano\meter}) to ease its detection. The largest xenon doping test ever performed in a LAr-TPC was carried out in ProtoDUNE-SP. From February to May 2020, a gradually increasing amount of xenon was injected to also compensate for the light loss due to air contamination. The response of such a large TPC has been studied using the ProtoDUNE-SP Photon Detection System (PDS) and a dedicated setup installed before the run. With the first it was possible to study the light collection efficiency with respect to the track position, while with the second it was possible to distinguish the xenon light (\SI{178}{\nano\meter}) from the LAr light (\SI{128}{\nano\meter}). 
The light shifting mechanism proved to be highly efficient even at small xenon concentrations ($<\SI{20}{\ppm}$ in mass) furthermore it allowed recovering the light quenched by pollutants. The light collection improved far from the detection plane, enhancing the photon detector response uniformity along the drift direction and confirming a longer Rayleigh scattering length for \SI{178}{\nano\meter} photons, with respect to \SI{128}{\nano\meter} ones. The charge collection by the TPC was monitored proving that xenon up to \SI{20}{\ppm} does not impact its performance.
}
\collaboration[c]{on behalf of the DUNE collaboration}
\proceeding{LIDINE2021 - Light Detection In Noble Elements\\
 14 -- 18 September 2021\\
University of California San Diego, San Diego, US}

\begin{document}
\maketitle
\flushbottom

\section{Introduction}
The Deep Underground Neutrino Experiment (DUNE) \cite{AbiI_2020} will be an international effort for the study of long-baseline oscillations of neutrinos produced at accelerators. It will perform precision measurements of oscillation parameters, investigate the neutrino mass hierarchy and the CP-violation (Charge Parity) in the leptonic sector. It will also explore physics without the use of the beam, being able to study supernova neutrino bursts (SNB) and to perform beyond the standard model (BSM) searches.

DUNE will use the LBNF accelerator facility for neutrino production, a near site with a set of detectors to monitor the beam and a far site, \SI{1300}{\kilo\meter} distant, with four detector modules with \SI{17}{\kilo\tonne} of liquid argon (LAr) each to reduce systematic uncertainties. Liquid Argon Time Projection Chamber (LAr-TPC) is the chosen solution for the DUNE far detector as it has become a mature technology for rare event physics \cite{ICARUS:2004wqc, MicroBooNE:2016pwy, Anderson:2012vc, article} ready to be scaled up to the multi-kiloton level.

Charged particles interacting with liquid argon release energy to the medium by ionization and excitation of argon atoms. Ionization electrons drift in an electric field towards anode wires where charge is collected, giving a 3D position and calorimetric information of the event. Liquid argon is also an excellent scintillator with a light yield about \SI{24000}{\ph/\mega\electronvolt} at the DUNE nominal electric field of \SI{500}{\volt/\centi\meter}. Scintillation light is peaked in the VUV (Vacuum UltraViolet) range at \SI{128}{\nano\meter} and emitted through a singlet ($\tau_f \sim \SI{7}{\nano\second}$) and a triplet ($\tau_s \sim \SI{1.3}{\micro\second}$) state de-excitation with a relative intensity depending on the particle stopping power. While the detection of VUV photons is challenging, advantages coming from it are fundamental for the DUNE experiment. It will allow improving the vertex reconstruction ($\SI{1}{\centi\meter} \xrightarrow{} \SI{1}{\milli\meter}$) and will give the absolute timing of the event. Furthermore, it is essential for non-beam physics as it will be the convenient method to trigger events in the case of SNB or BSM searches.


The DUNE Photon Detection System (PDS), based on the X-ARAPUCA technology \cite{AbiIV_2020}, can be enhanced by injecting xenon in argon at the level of a few \si{\ppm}.\footnote{In this paper, concentration quantities given in \si{\ppm} (part per million) are in terms of mass.} Xenon shifts part of the argon light, re-emitting photons at \SI{178}{\nano\meter}. Furthermore, the larger wavelength and the resulting lower refractive index translate into a higher Rayleigh scattering length \cite{argongroupspeed}, implying a potentially more uniform light collection efficiency throughout the detector drift volume. The light emission becomes faster in time allowing a better disentangling of events in the late light, furthermore it can also recover light that would otherwise be quenched by impurities. Previous studies in the literature were performed in small scale setups or in gas mixtures \cite{SUZUKI199367, Wahl:2014vma}, making this study the first ever implemented in a full scale LAr-TPC like the ProtoDUNE-SP detector.

\section{Xenon and nitrogen interaction mechanism with argon}
\begin{figure}
\begin{center}
\includegraphics[width=0.56\textwidth]{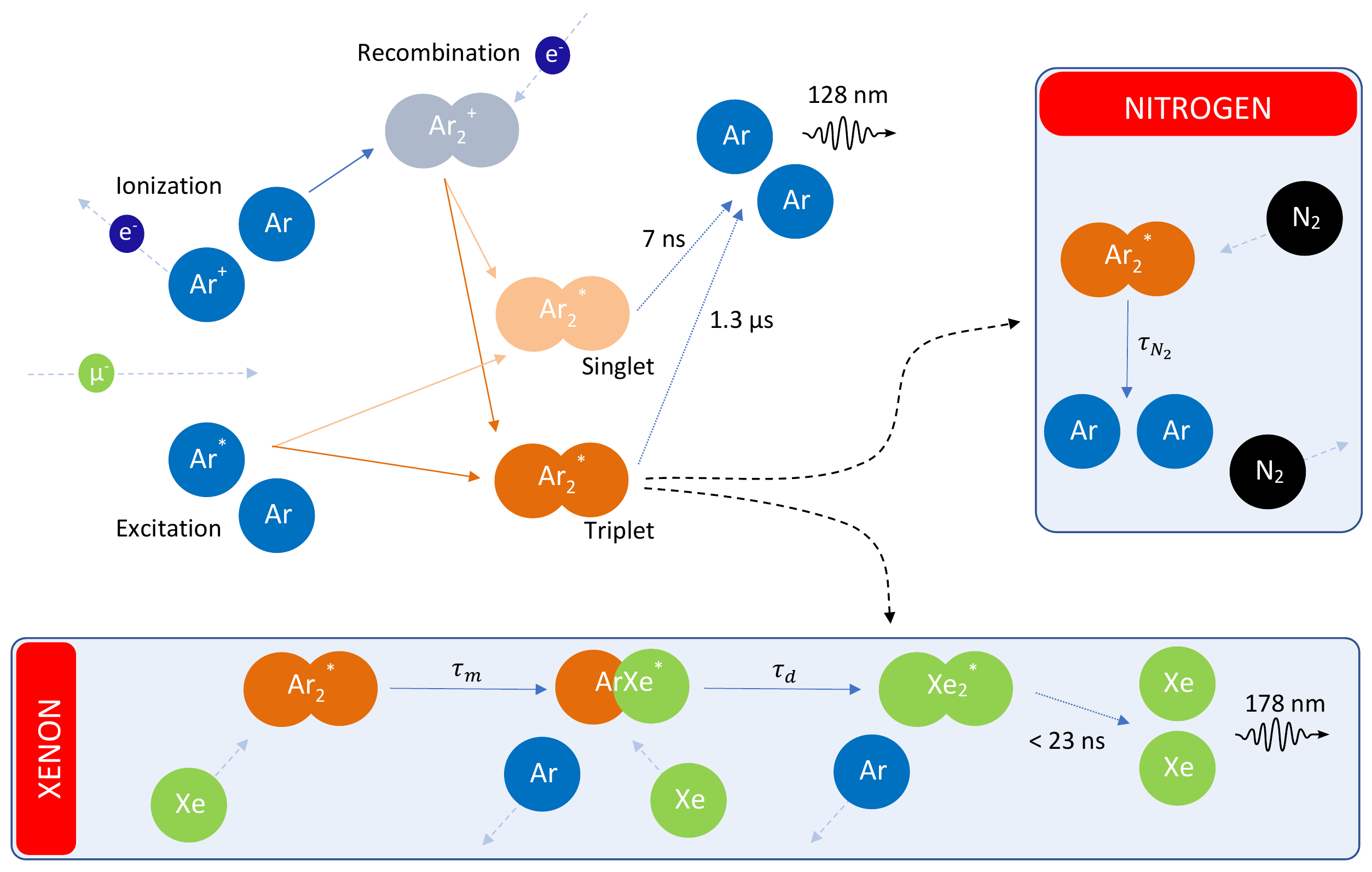}\hspace{6mm}%
\begin{minipage}[b]{0.4\textwidth}
    \caption{\label{fig:Xe-tranfer} On the left: scintillation model of liquid argon. On the bottom: energy transfer mechanism from excited argon state to xenon. The energy transfer time scales ($\tau_m$ and $\tau_d$) depend inversely on rate constants and xenon concentration, implying that the more xenon is injected, the more the mechanism is efficient until reaching saturation effects. On the right: non-radiative energy transfer from argon excimer to nitrogen that causes argon light quenching.}
    \vspace{-4mm}
\end{minipage}
\end{center}
\end{figure}
Ionizing particles (e.g. $e$, $\mu$, $\pi$) release their energy interacting with liquid argon (Figure~\ref{fig:Xe-tranfer}) that can be excited (Ar$^*$) or ionized (Ar$^+$) \cite{PhysRevD.103.043001}. In the first case the exciton forms an excited molecule Ar$_2^*$ interacting with argon atoms. In the second case the ion interacts first with argon, forming an ionized Ar$_2^+$ molecule that recombines with thermal electrons producing an Ar$_2^*$ excimer. The excited Ar$_2^*$ molecule can be formed in two states, a $^1\Sigma_u^+$ singlet and a $^3\Sigma_u^+$ triplet one that decay radiatively emitting \SI{128}{\nano\meter} photons. The de-excitation time constant is \SI{7}{\nano\second} and \SI{1.3}{\micro\second} for singlet and triplet states, respectively. If xenon is added to liquid argon, xenon atoms can interact with argon excimers creating an ArXe$^*$ excited dimer (Ar$_2^* + $ Xe $\xrightarrow[]{}$ ArXe$^*$ + Ar). Finally the produced dimer can come across a xenon atom resulting in the creation of Xe$_2^*$ that decays very fast ($<\SI{23}{\nano\second}$) emitting \SI{178}{\nano\meter} photons. The time constants of these processes $\tau_{m(d)} = [\text{Xe}]^{-1} k_{m(d)}^{-1}$ are inversely proportional to the rate constant \cite{Buzulutskov:2017wau} and to the xenon concentration, thus increasing the xenon concentration enhances the wavelength shifting efficiency until reaching saturation effects. The presence of pollutants like N$_2$ can reduce the light by quenching through collisional non radiative interaction with Ar$_2^*$ ($\text{Ar}_2^* + \text{N}_2 \xrightarrow{} 2 \text{Ar} + \text{N}_2$). This process is more efficient as the time scale becomes faster $\tau_{\text{N}_2} = [\text{N}_2]^{-1} k_{\text{N}_2}^{-1}$, i.e. at larger nitrogen concentrations. Xenon becomes a direct concurrent in the latter process and since its reaction rate constant is larger than the one of nitrogen, the light can be recovered with the wavelength shifting mechanism previously explained.

\section{Xenon doping procedure}
In ProtoDUNE-SP, xenon is injected in the gas phase of the argon recirculation system and far from the LAr condenser to allow the proper mixing before reaching the liquid phase. Small scale tests in a setup provided with a recirculation and purification system showed that an argon to xenon ratio for the gaseous mixture $\text{Ar}/\text{Xe}> 10^3$ is needed to avoid freeze-out effects. Considering liquid argon flow (CFD) simulations, the xenon injected is expected to be uniformly distributed inside the detector in few hours. 


Before the end of ProtoDUNE-SP Run 1, a sudden failure in the warm gas re-circulation pump leaked a certain amount of air inside the detector. Filters removed efficiently the oxygen but not the nitrogen. The xenon doping run was conducted entirely after this accident, and it was affected by N$_2$ contamination estimated to be around \SI{5.4}{ppm}.
Between February and May 2020 five xenon injections were performed in ProtoDUNE-SP and the response of the detector was monitored in the meanwhile. Each injection was about a few \si{\ppm} (1.1, 3.1, 7.4, 4.5, 2.8~\si{\ppm}), reaching a total amount of \SI{18.9}{\ppm} (\SI{13.5}{\kilo\gram}) at the end of the procedure.

\section{ProtoDUNE Single Phase}
\begin{figure}
    \centering
    \subfloat[]{
    \includegraphics[width=0.3\textwidth]{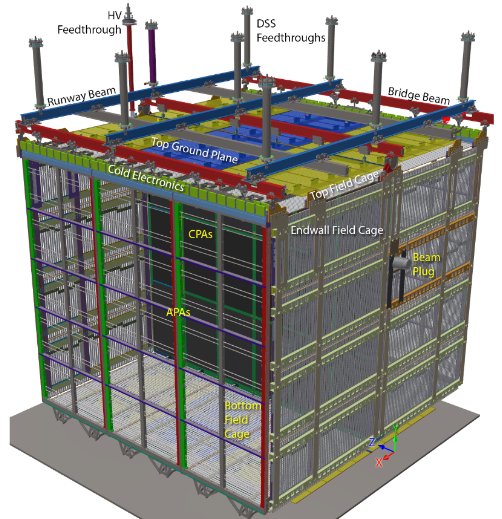}
    }
    \subfloat[]{
    \includegraphics[width=0.4\textwidth]{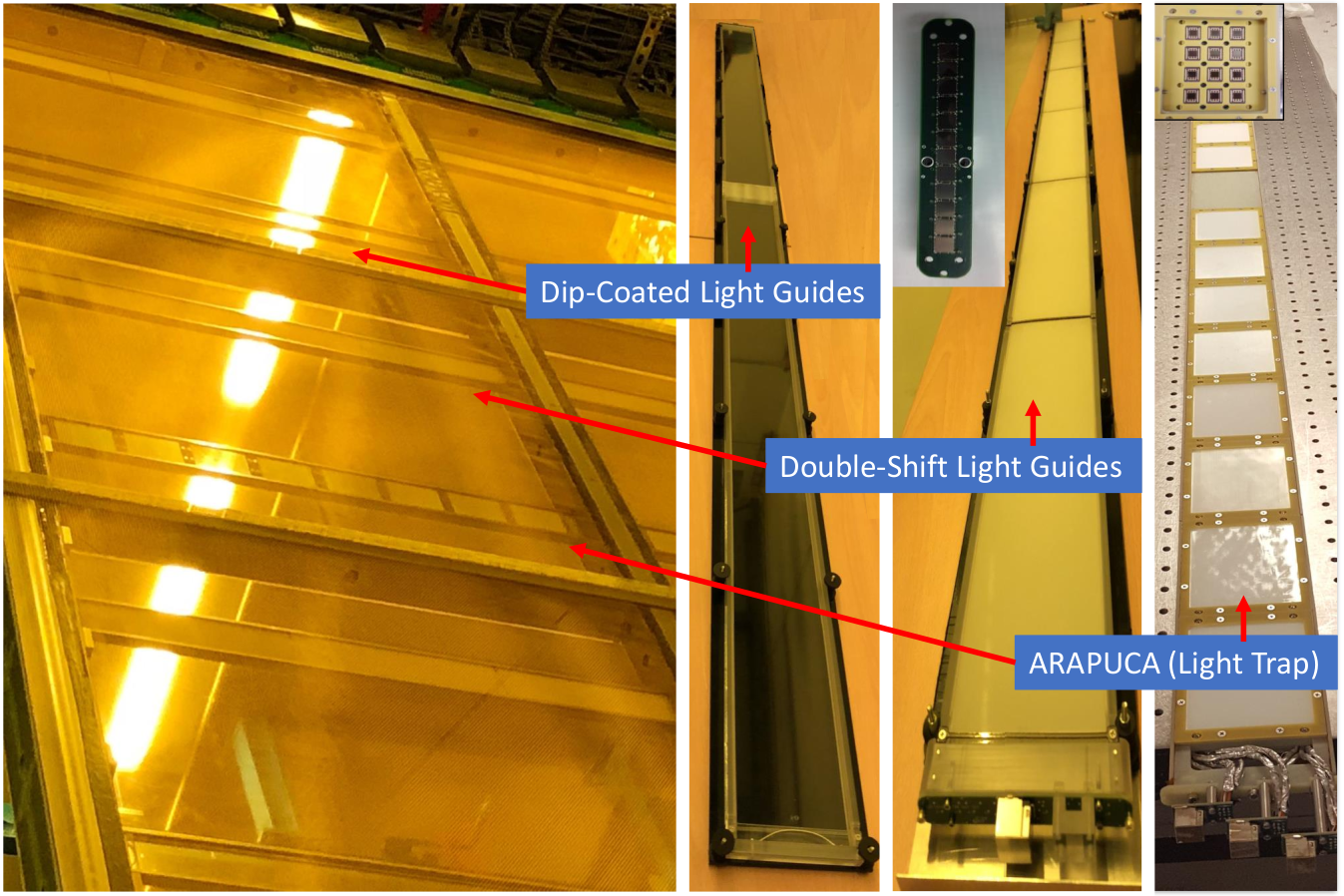}
    }
    \caption{(a) 3D view of ProtoDUNE-SP. (b) PDS modules installed inside an APA frame.}
    \label{fig:pdune}
\end{figure}

ProtoDUNE Single Phase \cite{dunecollaboration2021design} (Figure~\ref{fig:pdune}a) is the prototype of the first far detector module of DUNE. It is located at the CERN neutrino platform and served as test bench for LAr-TPC technologies devoted to future neutrino experiments. Its response has been studied during Run 1 with a beam of various particles ($\pi$, K, $\mu$, e) and with cosmic rays \cite{Abi_2020}. It contains \SI{770}{\tonne} of liquid argon  and the active volume is divided in two volumes by a Cathode Plane Assembly (CPA) that generates a \SI{500}{\volt/\centi\meter} drift electric field. The latter directs ionization electrons to the Anode Plane Assemblies (APA) where they are collected and readout.

\subsection{Photon Detection System}
Scintillation photons are detected with a Photon Detection System (PDS) composed by different modules (Figure~\ref{fig:pdune}b) that shift the \SI{128}{\nano\meter} light to a detectable wavelength for Silicon PhotoMultipliers (SiPM). 
Ten PDS modules are installed in each APA frame and equally spaced in the vertical direction. Three different technologies are implemented in ProtoDUNE-SP: double shift light guides \cite{HOWARD20189}, dip-coated light guides \cite{Moss_2015} and ARAPUCA. The ARAPUCA \cite{Machado_2016,Segreto_2018,Totani_2020} technology consist of a highly reflective box closed on all sides but one where a dichroic filter is located. The filter is externally coated with pTP (P-Terphenyl) shifting \SI{128}{\nano\meter} (or \SI{178}{\nano\meter}) light to \SI{350}{\nano\meter}, where it is transparent, and coated internally with TPB (Tetraphenyl Butadiene) that shifts the incoming light to \SI{430}{\nano\meter}, where the filter is reflective. Converted photons are then trapped and reflected inside the box until detected by SiPMs installed on the sides.

The PDS is triggered with the use of a Cosmic Ray Tagger (CRT) \cite{Abi_2020} that allows selecting horizontal crossing muons and provides the track position. It is located on two opposite sides of ProtoDUNE-SP, along the beam direction, and consists of scintillator strips ($\SI{5}{\centi\meter} \times \SI{365}{\centi\meter}$) arranged in two orthogonal layers, that cover a $\SI{6.8}{\meter} \times \SI{6.8}{\meter}$ surface on each side.

\subsection{Dedicated X-ARAPUCA setup}
\begin{figure}
    \centering
    \subfloat[]{
    \includegraphics[width=0.4\textwidth]{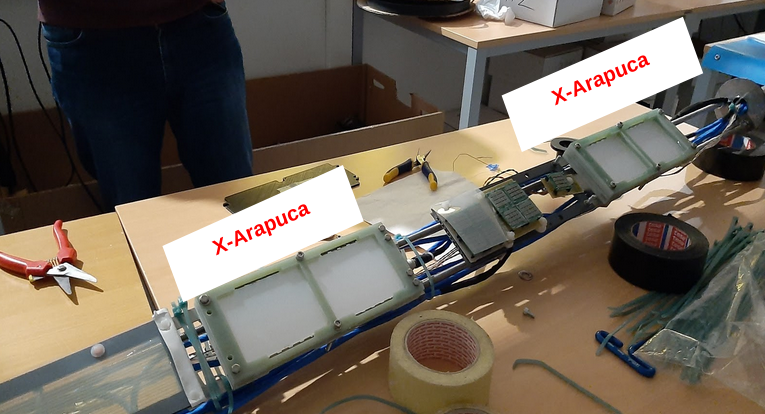}
    }
    \subfloat[]{
    \includegraphics[width=0.375\textwidth]{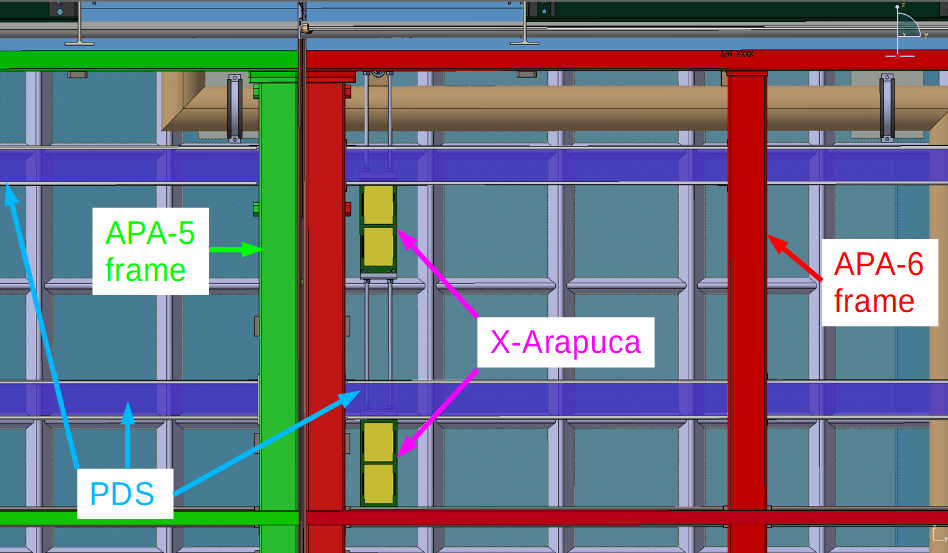}
    }
    \caption{(a) Dedicated setup with two X-ARAPUCA. (b) Sketch of the installation of the setup looking from the active volume through the APA and the cryostat wall.}
    \label{fig:xarapuca}
\end{figure}

To separate the light produced by xenon (\SI{178}{\nano\meter}) from the total light, a dedicated setup was inserted in ProtoDUNE-SP. It consists of two X-ARAPUCA  \cite{Machado_2018, Brizzolari_2021} devices (Figure~\ref{fig:xarapuca}a), an upgrade of the ARAPUCA concept where the TPB layer is replaced by a wavelength shifting guide, that were located behind one APA of ProtoDUNE-SP (Figure~\ref{fig:xarapuca}b). Since X-ARAPUCA are intrinsically sensitive to both \SI{128}{\nano\meter} and \SI{178}{\nano\meter} photons, a \SI{2}{\milli\meter} fused silica window was placed in front of the first module to stop argon scintillation light. In this paper, the first module is referred to as ``Xe XA'', as only sensitive to \SI{178}{\nano\meter} light, and the second one as ``Ar+Xe XA'', as it is sensitive to the total light.

The system was triggered by a standard triple coincidence of plastic scintillator bars placed on the top of ProtoDUNE-SP cryostat selecting vertical muons passing in front of the dedicated setup.

\section{Results}
\begin{figure}
    \centering
    \subfloat[]{
    \includegraphics[width=0.48\textwidth]{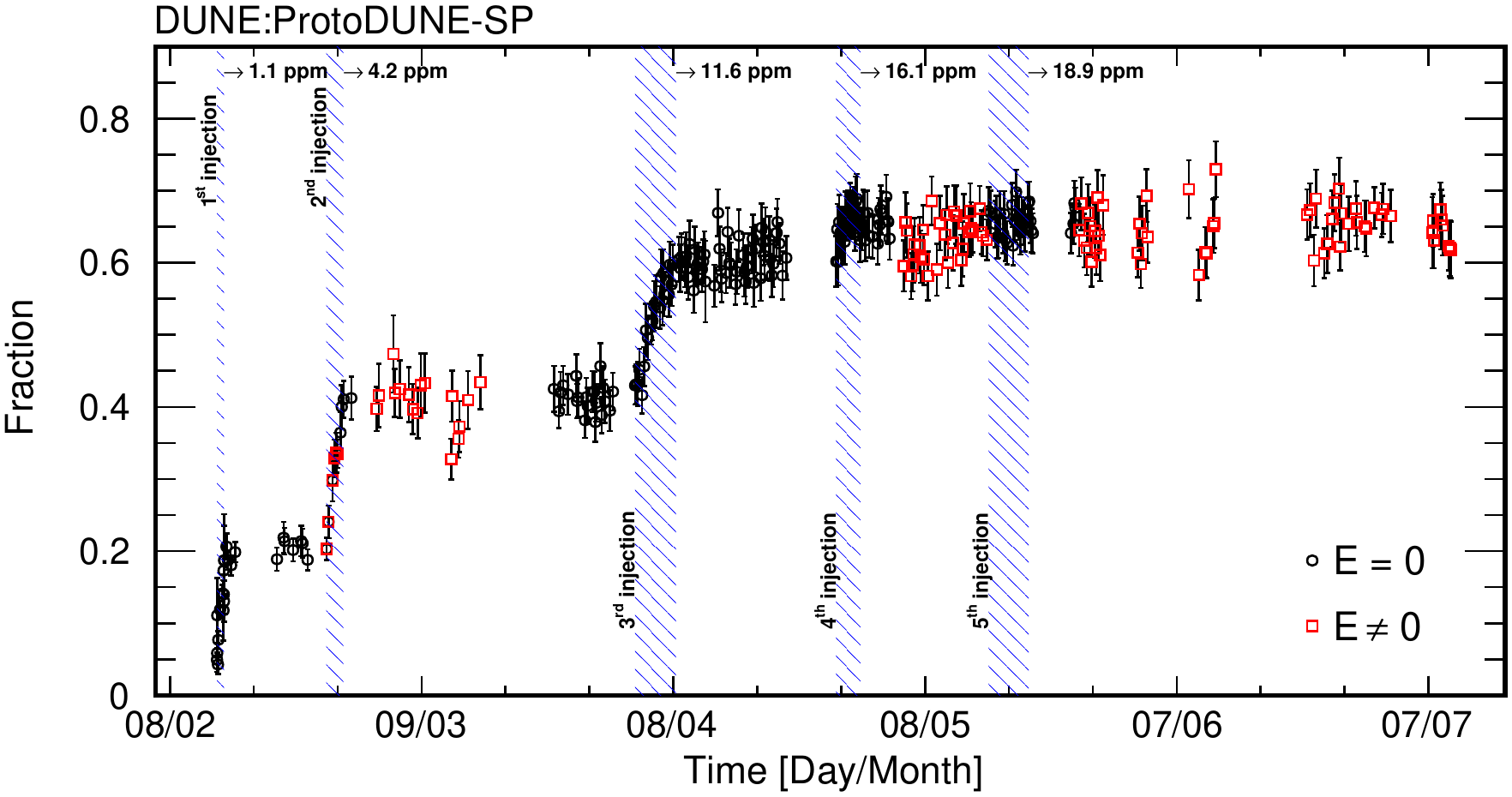}
    }
    ~
    \subfloat[]{
    \includegraphics[width=0.48\textwidth]{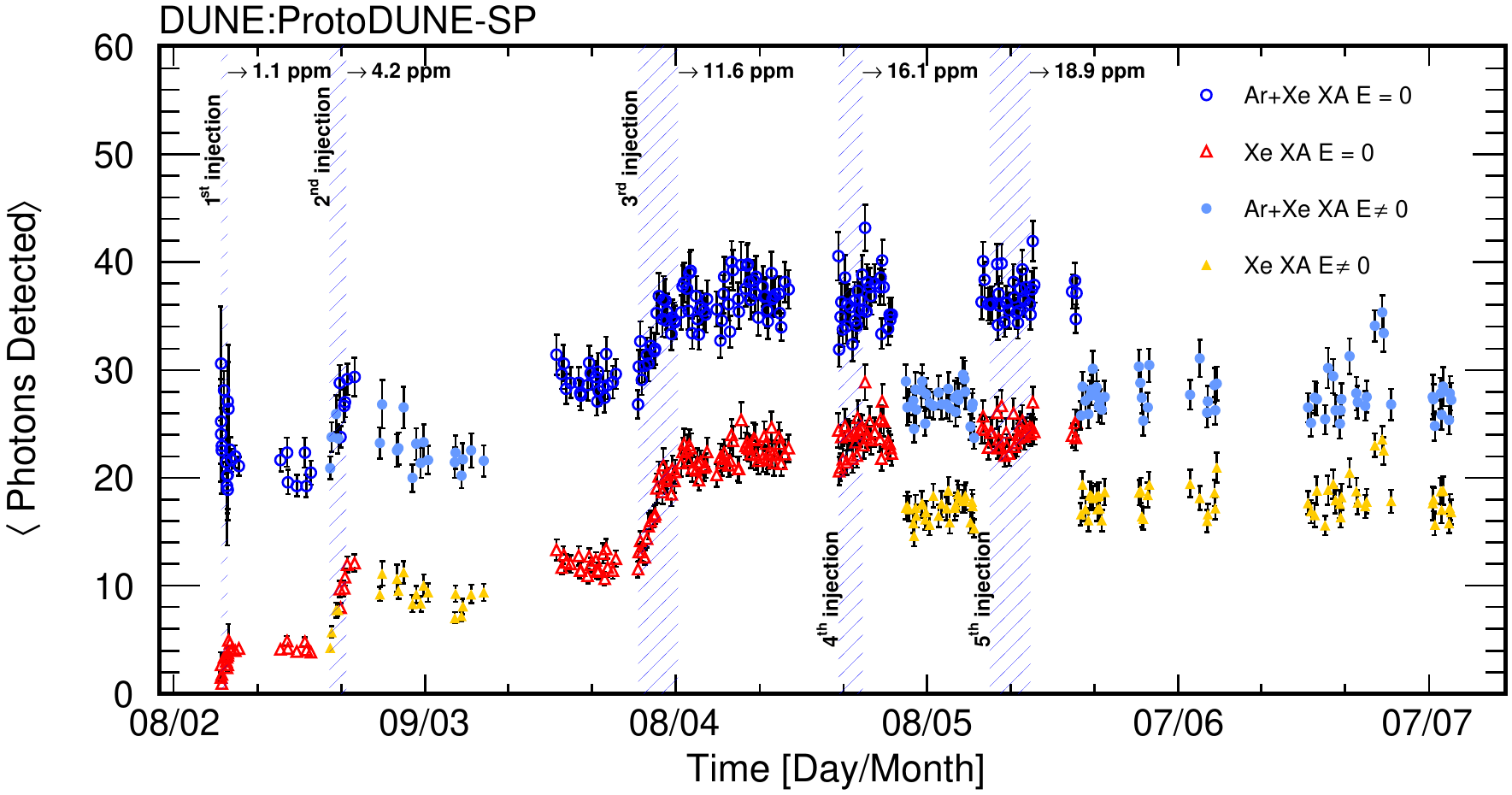}
    }
    \vspace{-3mm}
    
    \subfloat[]{
    \includegraphics[width=0.48\textwidth]{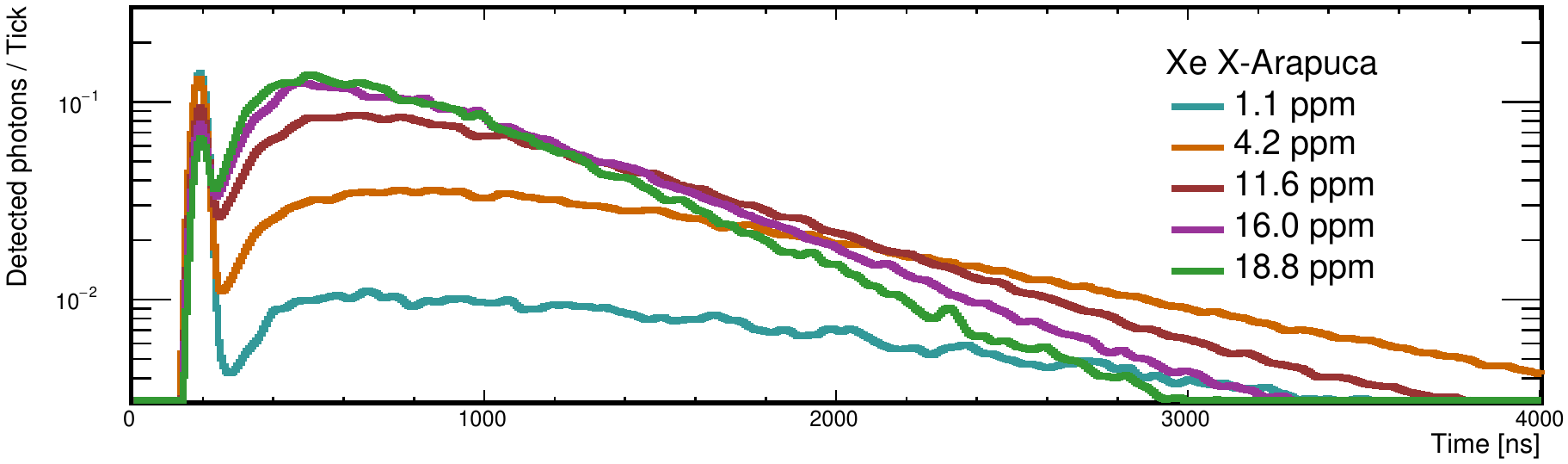}
    }
    ~
    \subfloat[]{
    \includegraphics[width=0.48\textwidth]{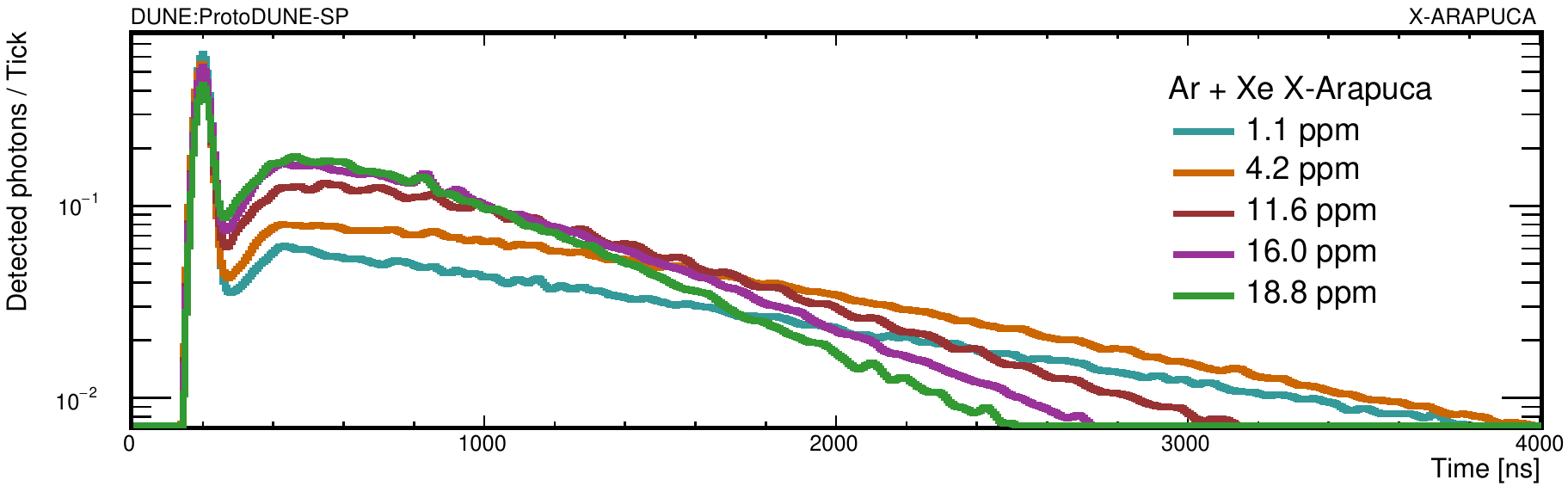}
    }
    \caption{(a): Fraction of wavelength-shifted light ($\SI{128}{\nano\meter} \xrightarrow[]{} \SI{178}{\nano\meter}$). (b) Average light detected by the dedicated X-ARAPUCA modules. Shaded areas show xenon injection periods. (c)(d) Average waveforms as a function of the xenon concentration for the ``Xe XA'' and the ``Ar+Xe XA'' respectively.}
    \label{fig:XA-result}
\end{figure}

The efficacy of the xenon doping mechanism can be tested measuring the amount of shifted light with respect to the total light detected. This observable, called Fraction (Eq.~\ref{eq:ratio}), is measured as the ratio of the average light detected by ``Xe XA'' with respect to the one of ``Ar+Xe XA''.
\begin{equation}
\text{Fraction} = \frac{\text{Xe light}}{\text{Ar light} + \text{Xe light}} \equiv \frac{\langle \gamma_{\text{Xe XA}} \rangle}{\langle \gamma_{\text{Ar+Xe XA}} \rangle}
\label{eq:ratio}
\end{equation}
The Fraction (Figure~\ref{fig:XA-result}a) increases with each doping and it flattens around \num{0.65} for xenon concentration greater than \SI{16.1}{ppm}. The detected light increases as well (Figure~\ref{fig:XA-result}b) as a function of xenon concentration, showing a recovery of the light otherwise quenched by N$_2$. Sudden drops in the detected light are due to the presence of the electric field and to the consequent reduction of the light coming from recombination.

The average waveform shape (Figure~\ref{fig:XA-result}c for xenon light detected by "Xe XA" module) changes with xenon injections, showing the typical bump shape, and becomes shorter in time for higher xenon concentrations. The time profile depends on the time constants related to the energy transfer processes; the rise time of the the bump is given by $\tau_{rise}^{-1} \sim \tau_{\text{N}_2}^{-1} + \tau_d^{-1} + \tau_{^3\Sigma_u^+}^{-1}$, while the decay time by $\tau_m$. The light detected by the "Ar+Xe XA" module (Figure~\ref{fig:XA-result}d) results in the sum of the xenon light and the fast argon component with a possible residual of argon slow component for low xenon concentrations.

\begin{figure}
    \centering
    \subfloat[]{
    \includegraphics[height=0.28\textwidth]{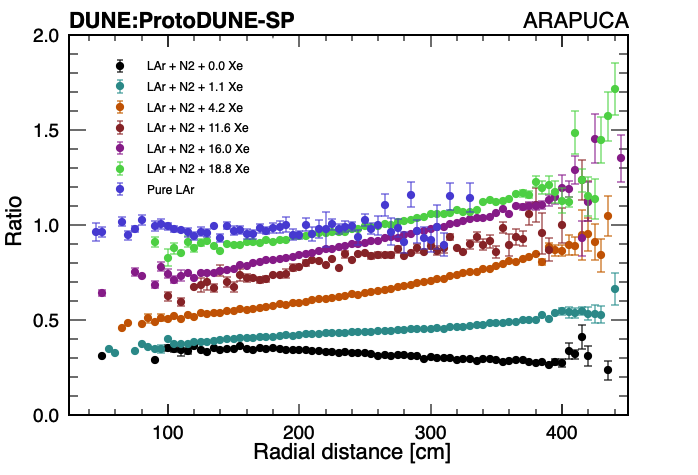}
    }
    \subfloat[]{
    \includegraphics[height=0.28\textwidth]{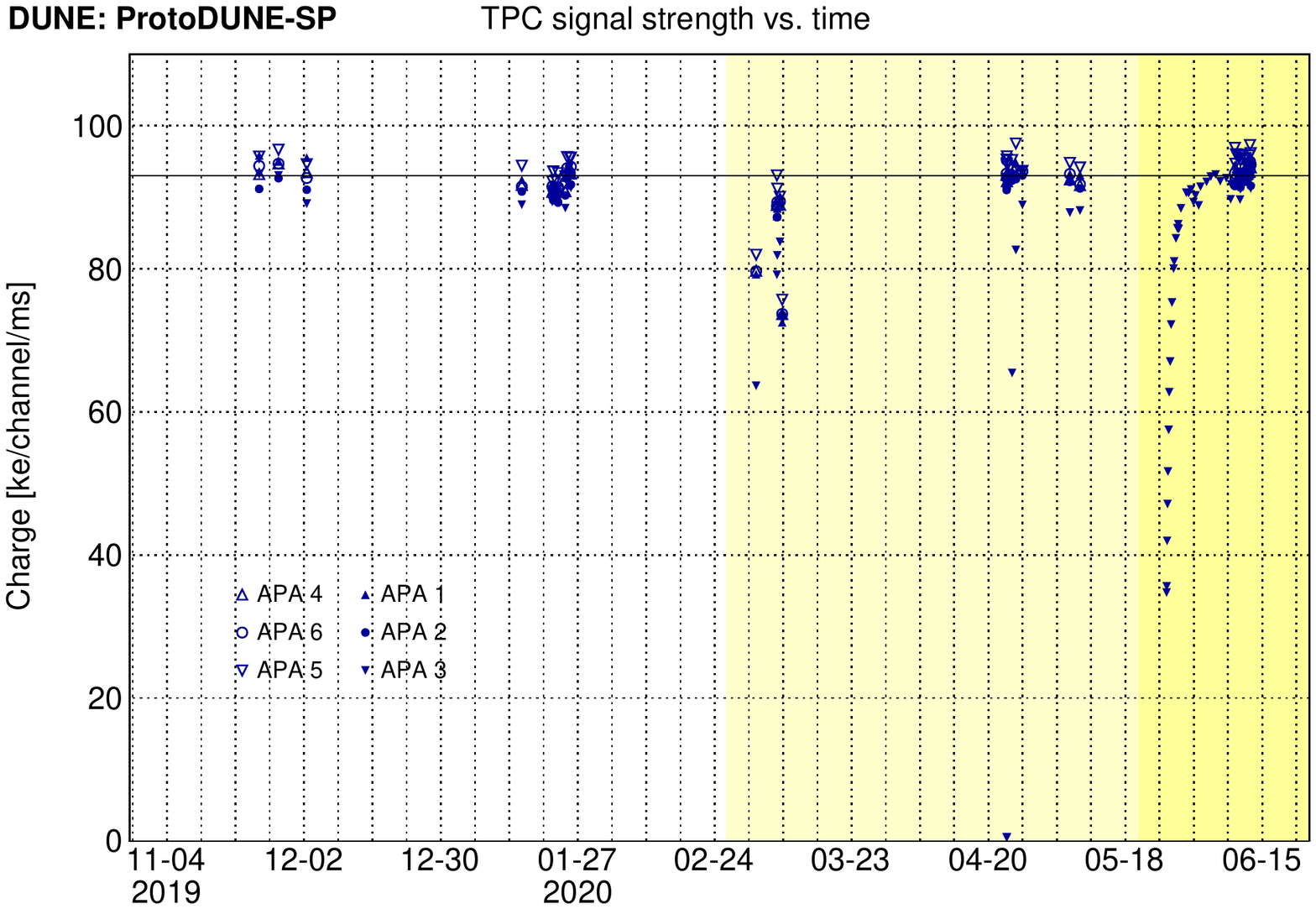}
    }
    \caption{(a) Ratio of collected light relative to the pure LAr (no $\text{N}_2$ contamination). (b) Charge signal strength readout by TPC before, during (light yellow) and after (yellow) the doping period. Line at \SI{93}{\kilo e$^-$/channel/\milli\second} is the typical reference for nominal voltages and high purity.}
    \label{fig:PDS-result}
\end{figure}

The response of the entire photon detection system can be understood using the ratio of the collected light relative to the pure liquid argon as a function of the track distance from the ProtoDUNE-SP photon detectors. The track distance is computed as the distance between PDS module center position and track point with same z coordinate (i.e. laying on the vertical plane perpendicular to the APA and passing through the PDS module center). This quantity (Figure~\ref{fig:PDS-result}a) shows that the accidental N$_2$ contamination reduced the scintillation light yield by more than a factor two, and the injection of xenon recovered the light loss. The light collection non-uniformity along the drift direction is given by Rayleigh scattering that reduces the collection efficiency for light emitted far from the readout point. As the slope of the curves increases with the xenon concentration, the collection efficiency far from the detection point increases, with respect to the pure argon case, implying a more uniform detector response.

To evaluate a possible impact of the xenon doping on the TPC, the charge collection efficiency was constantly monitored, using as figure of merit the signal strength parameter (Figure~\ref{fig:PDS-result}b). This is the average amount of charge readout on the TPC collection wires during a standard run with cosmic rays. Drops in signal strength on all APAs were only due to episodes where the purity dropped, while the response for APA-3 is always lower for the first few days after high voltage is turned on. Xenon injection did not cause any observable variation on charge collection.

\section{Conclusions}
In this paper, the first large scale test of xenon doping of liquid argon is presented: the TPC of ProtoDUNE-SP was successfully operated up to $\sim\SI{20}{\ppm}$ of xenon concentration. The response of the detector was studied with a dedicated setup based on the X-ARAPUCA technology and with the standard photon detection system of ProtoDUNE-SP. The light shifting mechanism of xenon ($\SI{128}{\nano\meter} \xrightarrow[]{} \SI{178}{\nano\meter}$) proved to be effective already at few \si{\ppm} level and tends to settle around \SI{16}{\ppm} of xenon injected. Furthermore the light loss due to the quenching of nitrogen is recovered by the injection of xenon. The study performed with respect to the track position showed an increased light collection efficiency far from the PDS detectors, compared to pure liquid argon; it results in an enhanced uniformity of the detector light collection and confirms a longer Rayleigh scattering length in liquid argon for \SI{178}{\nano\meter} light, with respect to \SI{128}{\nano\meter} photons. The charge collection of the TPC was tested against the xenon injection and no effect was observed up to $\sim\SI{20}{\ppm}$.

\clearpage
\bibliographystyle{JHEP}
\bibliography{biblio.bib}

\end{document}